\documentclass{aa}
\usepackage{graphicx}
\usepackage{subcaption}
\usepackage[colorlinks=true, citecolor=blue, linkcolor=blue, urlcolor=black]{hyperref}
\usepackage[varg]{txfonts}
\usepackage{appendix}
\usepackage{comment}
\usepackage{mathrsfs}
\usepackage{amsmath}
\usepackage{multicol}
\usepackage{multirow,tabularx}
\usepackage{longtable}
\DeclareUnicodeCharacter{2212}{-}
\newcommand{\RNum}[1]{\uppercase\expandafter{\romannumeral #1\relax}}

\usepackage[switch]{lineno}

\begin{document}
   \title{No strong radio absorption detected in the low-frequency spectra of radio-loud quasars at $z>5.6$}
   \author{A. J. Gloudemans \inst{\ref{inst1}}
   \and A. Saxena \inst{\ref{inst_ox}, \ref{inst_ucl}}
   \and H. Intema \inst{\ref{inst1}}
   \and J.~R.~Callingham \inst{\ref{inst_astron},\ref{inst1}} 
   \and K. J. Duncan \inst{\ref{inst_edin}}
   \and H. J. A. R\"{o}ttgering \inst{\ref{inst1}}
   \and S. Belladitta \inst{\ref{inst_heidel}, \ref{inst_inaf2}}
   \and M. J. Hardcastle \inst{\ref{inst_Hertfordshire}}
   \and Y. Harikane \inst{\ref{inst_tokyo}} 
   \and C. Spingola \inst{\ref{inst_inaf}} 
    }

   \institute{Leiden Observatory, Leiden University, PO Box 9513, 2300 RA Leiden, The Netherlands \\ e-mail: gloudemans@strw.leidenuniv.nl\label{inst1} \and  Department of Physics, University of Oxford, Denys Wilkinson Building, Keble Road, Oxford OX1 3RH, UK \label{inst_ox} \and Department of Physics and Astronomy, University College London, Gower Street, London WC1E 6BT, UK\label{inst_ucl} \and ASTRON, Netherlands Institute for Radio Astronomy, Oude Hoogeveensedijk 4, Dwingeloo, 7991 PD, The Netherlands\label{inst_astron} \and Institute for Astronomy, Royal Observatory, Blackford Hill, Edinburgh, EH9 3HJ, UK \label{inst_edin} \and Max Planck Institut f\"ur Astronomie, Königstuhl 17, D-69117, Heidelberg, Germany\label{inst_heidel} \and INAF — Osservatorio di Astrofisica e Scienza dello Spazio, via Gobetti 93/3, I-40129, Bologna, Italy \label{inst_inaf2} \and Centre for Astrophysics Research, University of Hertfordshire, Hatfield, AL10 9AB, UK \label{inst_Hertfordshire} \and Institute for Cosmic Ray Research, the University of Tokyo, 5-1-5 Kashiwa-no-Ha, Kashiwa City, Chiba, 277-8582, Japan \label{inst_tokyo} \and INAF – Istituto di Radioastronomia, Via Gobetti 101, 40129 Bologna, Italy \label{inst_inaf}}
   
   \date{Received: 27 July 2023 / Accepted: 23 August 2023}
 
 \abstract{We present the low-frequency radio spectra of 9 high-redshift quasars at $5.6 \leq z \leq 6.6$ using the Giant Metre Radio Telescope band-3, -4, and -5 observations ($\sim$300-1200 MHz), archival Low Frequency Array (LOFAR; 144 MHz), and Very Large Array (VLA; 1.4 and 3 GHz) data. Five of the quasars in our sample have been discovered recently, representing some of the highest redshift radio bright quasars known at low-frequencies. We model their radio spectra to study their radio emission mechanism and age of the radio jets by constraining the spectral turnover caused by synchrotron self-absorption (SSA) or free-free absorption (FFA). Besides J0309+2717, a blazar at $z=6.1$, our quasars show no sign of a spectral flattening between 144 MHz and a few GHz, indicating there is no strong SSA or FFA absorption in the observed frequency range. However, we find a wide range of spectral indices between $-1.6$ and $0.05$, including the discovery of 3 potential ultra-steep spectrum quasars. Using further archival VLBA data, we confirm that the radio SED of the blazar J0309+2717 likely turns over at a rest-frame frequency of 0.6-2.3 GHz (90-330 MHz observed frame), with a high-frequency break indicative of radiative ageing of the electron population in the radio lobes. Ultra-low frequency data below 50 MHz are necessary to constrain the absorption mechanism for J0309+2717 and the turnover frequencies for the other high-$z$ quasars in our sample.
A relation between linear radio jet size and turnover frequency has been established at low redshifts. If this relation were to hold at high redshifts, the limits on the turnover frequency of our sample suggest the radio jet sizes must be more extended than the typical sizes observed in other radio-bright quasars at similar redshift. To confirm this deep radio follow-up observations with high spatial resolution are required. }


\maketitle

\section{Introduction}
\label{sec:introduction}

High-redshift ($z>5.6$) quasars are useful probes of early supermassive black hole (SMBH) formation and evolution, as well as the Epoch of Reionization (EoR), when the Universe changed from neutral to ionised. Currently, hundreds of quasars have been discovered at $z>6$, with the highest known redshift quasar at $z=7.6$ \citep{wang2021ApJ...907L...1W}. Previous work has shown that about 10\% of the quasars at $z>5$ can be classified as radio-loud (RL; e.g. \citealt{Banados2015ApJ...804..118B, Gloudemans2021A&A...656A.137G}), which is defined as $R=f_{5\text{GHz}}/f_{4400\text{\AA}} > 10$, where $f_{5\text{GHz}}$ and $f_{4400\text{\AA}}$ are the flux density at 5 GHz and 4400 $\text{\AA}$ rest-frame, respectively. These RL quasars at $z \ge 5.5$ are extremely rare as only about a dozen objects are known \citep{McGreer2006ApJ...652..157M, Willott2010AJ....139..906W, Belladitta2020A&A...635L...7B, Banados2021ApJ...909...80B, Endsley2022arXiv220600018E, Gloudemans2022A&A...668A..27G}. They are highly interesting objects to study SMBH activity in the early Universe and can be used to measure the 21-cm absorption line to directly determine the neutral hydrogen content in the EoR (see e.g. \citealt{carilli2002ApJ...577...22C, Mack2012MNRAS.425.2988M, ciardi2015aska.confE...6C}).

Detailed studies of the radio spectra of powerful radio quasars at high redshifts can help constrain the physical mechanisms by which quasars are triggered and the lifetimes of SMBH activity through determination of the ages of their radio jets (e.g. \citealt{Saxena2017MNRAS.469.4083S, Shao2022A&A...659A.159S}). The origin and evolution of the radio activity of their active galactic nuclei (AGN) and its impact on the host galaxy via AGN `feedback' are outstanding problems in the field, and can be studied by observing young and compact radio sources with their radio lobes situated within the optical host galaxy. 

Radio sources with small projected linear sizes ($\lesssim$ 20 kpc) and a peak in their broadband radio spectra are classified as peaked-spectrum (PS) sources, with subclasses of compact steep spectrum (CSS), gigahertz-peaked spectrum (GPS), and high-frequency peaked (HFP) sources \citep[e.g.][]{fanti1990A&A...231..333F, Odea1991ApJ...380...66O, Dallacasa2000A&A...363..887D, Stanghellini2009AN....330..223S, Odea2021A&ARv..29....3O}. There are two hypotheses for the small linear size and spectral shape of PS sources: the `youth' hypothesis and the `frustration' hypothesis. The youth hypothesis suggests these PS sources are young radio galaxies that will grow into large scale radio-loud AGN while they subsequently evolve through a HFP, GPS, and CSS phase \citep[e.g.][]{Slob2022A&A...668A.186S}. The frustration hypothesis suggests PS sources have small linear sizes because they are situated in dense environments that slow down the growth of the radio jets. Observations have provided evidence for both hypothesis by showing a strong connection to their environment (e.g. \citealt{Wilkinson1984Natur.308..619W, Tingay2003AJ....126..723T, Stanghellini2009AN....330..153S, Kunert2010MNRAS.408.2261K, callingham2015ApJ...809..168C, Saxena2017MNRAS.469.4083S}) and small double-lobed structures with a relation between their linear scale and both radio power and turnover frequency (e.g. \citealt{Odea1997AJ....113..148O, Snellen2000MNRAS.319..445S, Kunert2010MNRAS.408.2261K}). The physical origin of the turnover in the spectrum of PS sources could be due to a free-free absorption (FFA) or synchrotron self-absorption (SSA) mechanism, or a combination of them. 

Most RL quasar discoveries at high-redshift have been made using radio surveys at high frequencies (1-10 GHz; e.g. \citealt{McGreer2006ApJ...652..157M, Banados2015ApJ...804..118B, Belladitta2020A&A...635L...7B}). However, since their radio spectra are redshifted significantly towards the low frequency regime at $z>5$, observations at low-frequencies are vital to constrain their spectral shapes. The unprecedented area coverage and sensitivity (in the $\sim 100\mu$Jy regime) of the Low Frequency Array (LOFAR; \citealt{vanHaarlem2013A&A...556A...2V}) now enables the detection of large numbers of high-$z$ quasars at frequencies of 50-200 MHz. Specifically, the LOFAR Two Metre Sky Survey (LoTSS-DR2; \citealt{Shimwell2022A&A...659A...1S}) covers 5720 deg$^2$ and reaches such a depth (median noise level $\sim83$ $\mu$Jy beam$^{-1}$) that 36\% of the known quasar population at $z>5$ have been detected at $>2\sigma$ significance \citep{Gloudemans2021A&A...656A.137G}. Furthermore, \cite{Gloudemans2022A&A...668A..27G} utilised this survey, together with the DESI Legacy Imaging Survey \citep{dey2019AJ....157..168D} to discover 24 new radio-bright quasars at $4.9 \leq z \leq6.6$. The recent discovery of new high-$z$ quasars allows for detailed investigation of their radio spectral shapes with the goal of constraining the ages of the radio jets and radio emission mechanisms. 

In this work, we study the low-frequency radio spectra of a sample of 9 quasars at $5.6 \leq z \leq6.6$ (5 of which have been recently discovered by \citealt{Gloudemans2022A&A...668A..27G}) using the Giant Metrewave Radio Telescope (GMRT; \citealt{Swarup1991ASPC...19..376S}) and archival data from both LOFAR, the Very Large Array (VLA; \citealt{Thompson1980ApJS...44..151T}), and Australian SKA Pathfinder (ASKAP; \citealt{Johnston2008ExA....22..151J}). Some previous studies have investigated the broad-band radio spectra of quasars at high-redshift. For example, \cite{Coppejans2017MNRAS.467.2039C} investigated the radio spectra of 30 quasars at $z>4.5$ (including four quasars at $z>5.6$) and found diverse radio spectra, despite certain selection effects, with 34\% classified as PS sources. A study by \cite{Shao2022A&A...659A.159S} presented the spectral turnover of 9 radio-loud quasars at $ 5.0 \leq z \leq 6.1$ and found turnover frequencies between $\sim$1-50 GHz in the rest-frame, thereby classifying them as GPS and HFP sources. One of the currently highest known redshift radio-loud quasar PSO J172.3556+18.7734 or J1129+1846 (hereafter J172+18) at $z=6.8$ \citep{Banados2021ApJ...909...80B} has also been suggested to show a turnover at a few GHz in the rest-frame by \cite{Momjian2021AJ....161..207M}, with further evidence for a turnover provided by \cite{Gloudemans2021A&A...656A.137G} with a LOFAR non-detection. Our sample complements these previous works as it probes quasars out to higher redshifts and is selected to be bright at low-frequencies, which is the important frequency for 21cm science, since the 21cm line is shifted into the MHz regime at $z>6$. Furthermore, building a statistical sample of radio spectral measurements of these rare sources will be important to determine the formation and evolution of the first population of radio-loud AGN. 

This paper is structured as follows. In Section~\ref{sec:obs}, we describe the sample selection, observations, and data reduction process. In Section~\ref{sec:modelling}, we outline the different absorption models that we used to fit our spectra and our modelling procedure. In Section~\ref{sec:results}, we present the resulting radio spectra of our sample and in Section~\ref{subsec:banados_quasar} that of the quasar J172+18 at $z=6.8$. We constrain the linear sizes of our sample and compare them to the literature in Section~\ref{sec:discussion}. Finally, in Section~\ref{sec:conclusion} the conclusions are presented. In this paper, we use the AB magnitude system \citep{Oke1983ApJ...266..713O} and assume a $\Lambda$-CDM cosmology with H$_{0}$= 70 km s$^{-1}$ Mpc$^{-1}$, $\Omega_{M}$ = 0.3, and $\Omega_{\Lambda}$ = 0.7.

\section{Observations \& data reduction}
\label{sec:obs}

\begin{figure}
    \centering
    \includegraphics[width=1.0\columnwidth]{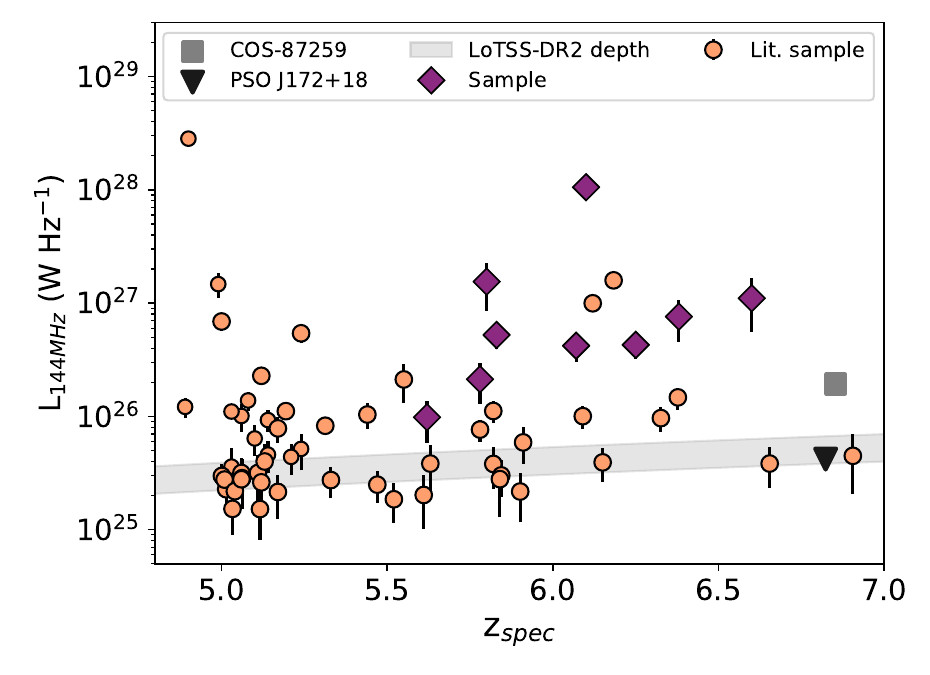}
    \caption{Radio luminosities at 144 MHz versus spectroscopic redshift of our selected sample and other known LOFAR detected quasars compiled by \cite{Gloudemans2021A&A...656A.137G, Gloudemans2022A&A...668A..27G}. Our sample probes some of the brightest and highest redshift radio-loud quasars currently known at low frequencies. The heavily obscured AGN COS-87259 is currently the highest redshift radio-loud quasar known and has been detected by LOFAR \citep{Endsley2022arXiv220201219E, Endsley2022arXiv220600018E}. The two radio-loud quasars around $z\sim6.1$ (J1427+3312 and J1429+5447) have not been included in our sample because their radio spectra have already been studied by \cite{Shao2020A&A...641A..85S} and \cite{Shao2022A&A...659A.159S}, with J1427+3312 showing a turnover around 1.7 GHz rest-frame. }
    \label{fig:radio_lum}
\end{figure}

To constrain the radio spectra of these LOFAR detected high-$z$ quasars, we utilise GMRT, which operates at six frequency bands ranging from $\sim$150-1500 MHz. In this section, we describe our sample selection, the GMRT observations we have conducted, and the ancillary radio data used. 

\subsection{Sample selection}
\label{subsec:sample_selection}

Our sample consists of 9 high-$z$ quasars selected from multiple works in the literature (see Table~\ref{tab:obs_details}) with spectroscopic redshifts ranging from 5.6 to 6.6. The majority of our targets (5/9) were recently discovered in LoTSS by \cite{Gloudemans2022A&A...668A..27G}. The target selection is based on their high low-frequency radio luminosities and redshifts, as illustrated in Fig.~\ref{fig:radio_lum}, and their visibility with GMRT. The new radio-loud high-$z$ quasars discovered by \cite{Gloudemans2022A&A...668A..27G} and other works allow for investigation of the radio spectra of the highest redshift and radio-brightest quasars to date. The targets are summarised in Table \ref{tab:obs_details}. The two radio-loud quasars around $z\sim6.1$ (J1427+3312 and J1429+5447) have not been included because their radio spectra have been studied already by \cite{Shao2020A&A...641A..85S} and \cite{Shao2022A&A...659A.159S}. In their work J1427+3312 shows a turnover around 1.7 GHz rest-frame and J1429+5447 shows no turnover above $\sim1$ GHz with $\alpha=-0.67^{+0.04}_{-0.03}$. These and other known quasars will be further discussed in Sect. \ref{subsec:discuss_linear_sizes}. The radio spectrum of J0309+2717, a blazar discovered by \cite{Belladitta2020A&A...635L...7B} at $z=6.1$, has been studied in \cite{Spingola2020A&A...643L..12S} using arcsec angular resolution observations with the Very Large Array (VLA) and milliarcsec angular resolution with the Very Long Baseline Array (VLBA). However, together with data from the TIFR GMRT Sky Survey (TGSS; \citealt{Intema2017A&A...598A..78I}) at 147 MHz there is still an unexplored frequency range between 0.15 and 1.4 GHz, where there could be a turnover, and therefore we included this source in our sample.

\begin{table*}
\caption{Details of our quasar sample and the GMRT observations. The RMS noise level is given for the full band integration. }
\label{tab:obs_details}      
\centering
\resizebox{0.95\textwidth}{!}{
\begin{tabular}{c c c c c c c c c c}  
\hline\hline
Source name & Optical coordinates & Redshift & Ref. & Band & Phase cal. & Flux cal. & Exp. time & Beam size & RMS \\ 
 & (J2000) &  & & & & & (mins) & (arcsec$^{2}$) & (mJy beam$^{-1}$) \\
\hline
J0002+2550 & 00:02:39.39 +25:50:34.96 & 5.80 $\pm$ 0.02 & 1 & 3 & 0040+331 & 3C48 & 101 & 6.5 $\times$ 7.6 & 0.04\\ 
 &  & & & 4 & 0040+331 & 3C286 & 96 & 3.1 $\times$ 7.4 & 0.02\\ 
J0309+2717 & 03:09:47.49 +27:17:57.31 & 6.10 $\pm$ 0.03 & 2 & 3 & 0318+164 & 3C48 & 101 & 5.4 $\times$ 8.0 & 0.06\\ 
 &  & & & 4 & 0318+164 & 3C48 & 81 & 3.2 $\times$ 4.1 & 0.02 \\ 
J0803+3138 & 08:03:05.42 +31:38:34.20 & 6.384 $\pm$ 0.004 & 3 & 3 & 0744+378 & 3C48 & 121 & 5.5 $\times$ 13.8 & 0.04\\ 
 &  & & & 4 & 0744+378 & 3C48 & 81 & 3.0 $\times$ 7.7 & 0.02\\ 
  &  & & & 5 & 0735+331 & 3C147 & 101 & 1.8 $\times$ 3.0 & 0.02\\ 
J0912+6658 & 09:12:07.67 +66:58:46.45 & 5.62 $\pm$ 0.02 & 4 & 3 & 0834+555 & 3C48 & 121 & 7.2 $\times$ 17.4 & 0.08\\ 
 &  & & & 4 & 0834+555 & 3C48 & 112 & 3.1 $\times$ 12.3 & 0.02 \\  
  &  & & & 5 & 0834+555 & 3C147 & 101 & 1.7 $\times$ 4.5 & 0.02\\ 
J1037+4033 & 10:37:58.17 +40:33:29.08 & 6.07 $\pm$ 0.03 & 4 & 3 & 1006+349 & 3C48 & 101 & 5.1 $\times$ 14.1 & 0.08\\ 
 &  & & & 4 & 1006+349 & 3C48 & 81 & 2.8 $\times$ 9.1 & 0.03 \\ 
J1133+4814 & 11:33:50.41 +48:14:31.35 & 6.25 $\pm$ 0.02 & 4 & 3 & 1145+497 & 3C48 & 70 & 5.1 $\times$ 14.9 & 0.05\\ 
 &  & & & 4 & 1145+497 & 3C48 & 81 & 2.9 $\times$ 8.8 & 0.02 \\ 
J1545+6028 & 15:45:52.09  +60:28:23.95 & 5.78 $\pm$ 0.03 & 5 & 3 & 1449+632 & 3C48 & 121 & 4.5 $\times$ 9.3 & 0.05\\ 
 &  & & & 4 & 1449+632 & 3C286 & 118 & 3.3 $\times$ 5.2 & 0.02\\ 
J2201+2338 & 22:01:07.62 +23:38:37.87 & 5.83 $\pm$ 0.02 & 4 & 3 & 2251+188 & 3C48 & 141 & 5.3 $\times$ 12.2 & 0.04\\ 
 &  & &  & 4 & 2251+188 & 3C286 & 100 & 3.2 $\times$ 7.0 & 0.02\\ 
  &  & & & 5 & 2251+188 & 3C48 & 45 & 1.8 $\times$ 3.6 & 0.02\\ 
J2336+1842 & 23:36:24.72 +18:42:47.98 & 6.6 $\pm$ 0.04 & 4 & 3 & 2251+188 & 3C48 & 141 & 5.4 $\times$ 12.0 & 0.04\\ 
 &  & & & 4 & 2251+188 & 3C286 & 101 & 3.2 $\times$ 8.8 & 0.03\\ 
  &  & & & 5 & 2251+188 & 3C48 & 45 & 1.8 $\times$ 2.9 & 0.03\\ 
\hline \hline
\end{tabular}}
\vspace{0.2cm} \raggedright { \small \textbf{References.} [1] \cite{Fan2004AJ....128..515F} [2] \cite{Belladitta2020A&A...635L...7B} [3] \cite{Wang2019ApJ...884...30W} [4] \cite{Gloudemans2022A&A...668A..27G} [5] \cite{WangF2016} } 
\end{table*}

\subsection{uGMRT observations and data reduction}

The observations of our sample were conducted using the upgraded GMRT (uGMRT; see \citealt{Gupta2017CSci..113..707G}) in August 2022 (proposal ID 42$\textunderscore$037; PI Gloudemans). All 9 targets were observed in band 3 (250-500 MHz) and band 4 (550-900 MHz), and 4 targets that were not detected by other surveys at GHz frequency were also observed in band 5 (1000-1450 MHz). The details of the observations are summarised in Table \ref{tab:obs_details}. Each observing block started with observing one of the flux density calibrators (3C48/3C147/3C286) for $\sim20$ minutes, followed by a sequence of phase calibrator, target observation, and again phase calibrator. We observed the phase calibrators (listed in Table ~\ref{tab:obs_details}) for about 5 minutes during each sequence and observed the targets between 45-141 minutes depending on the frequency band and radio brightness of the source. The theoretical beam sizes are 9$\arcsec$, 5$\arcsec$, and  2$\arcsec$ for band 3, 4, and 5, respectively (see \citealt{Swarup1991ASPC...19..376S}).

\begin{table*}
\caption{Measured integrated flux densities from GMRT observations and the LoTSS-DR2, RACS, FIRST, and VLASS surveys (with added 10\% uncertainty on the flux density errors). All integrated flux densities are in units of mJy. The spectral index $\alpha$ is determined from the power-law fitting routine. The literature references for the UV magnitudes M$_{1450\AA}$ of each source are indicated in the final column. } 
\label{tab:measurements}      
\centering
\resizebox{0.99\textwidth}{!}{
\begin{tabular}{c c c c c c c c c c c c c c c} 
\hline\hline
Source name & Redshift & $\alpha$ & S$_{144\text{MHz}}$ & S$_{383\text{MHz}}$ & S$_{675\text{MHz}}$ & S$_{1260\text{MHz}}$ & S$_{887\text{MHz}}$ & S$_{1.4\text{GHz}}$ &  S$_{3\text{GHz}}$ & M$_{1450\AA}$ & Ref. \\ 
\hline

J0002+2550 & 5.80$\pm$0.02 & $-1.6^{+0.19}_{-0.12}$ & 1.33$\pm$0.37 & 0.33$\pm$0.1 & 0.15$\pm$0.05 & -  & -  & -  & - & $-27.66$ & 1 \\[2pt]
J0309+2717 & 6.10$\pm$0.03 & $-0.51^{+0.04}_{-0.05}$ & 66.5$\pm$7.1 & 40.8$\pm$4.3 & 37.7$\pm$3.9 & -  & 35.6$\pm$4.0 & 23.9$\pm$3.3 & 13.7$\pm$1.6 & $-25.1$ & 2 \\[2pt]
J0803+3138 & 6.38$\pm$0.0 & $-1.12^{+0.16}_{-0.12}$ & 1.3$\pm$0.32 & 0.38$\pm$0.1 & 0.3$\pm$0.06 & 0.14$\pm$0.05 & -  & -  & - & $-26.49$ & 3 \\[2pt]
J0912+6658 & 5.62$\pm$0.02 & $-0.23^{+0.16}_{-0.17}$ & 1.23$\pm$0.26 & 0.47$\pm$0.17 & 0.63$\pm$0.1 & 0.79$\pm$0.12 & -  & -  & 0.9$\pm$0.34 & $-26.43$ & 4 \\[2pt]
J1037+4033 & 6.07$\pm$0.03 & $0.06^{+0.12}_{-0.13}$ & 8.16$\pm$0.97 & 4.94$\pm$0.63 & 6.17$\pm$0.67 & -  & 9.28$\pm$1.42 & 9.42$\pm$1.09 & 10.5$\pm$1.3 & $-25.25$ & 4 \\[2pt]
J1133+4814 & 6.25$\pm$0.02 & $-0.21^{+0.09}_{-0.08}$ & 4.67$\pm$0.79 & 3.48$\pm$0.44 & 2.79$\pm$0.31 & -  & -  & 3.23$\pm$0.46 & 2.69$\pm$0.63 & $-25.0$ & 4 \\[2pt]
J1545+6028 & 5.78$\pm$0.03 & $-0.79^{+0.16}_{-0.1}$ & 0.87$\pm$0.24 & 0.48$\pm$0.17 & 0.29$\pm$0.07 & -  & -  & -  & - & $-27.37$ & 5 \\[2pt]
J2201+2338 & 5.83$\pm$0.02 & $-0.51^{+0.07}_{-0.06}$ & 3.6$\pm$0.73 & 2.32$\pm$0.3 & 2.23$\pm$0.26 & 1.33$\pm$0.17 & 1.78$\pm$0.55 & -  & 0.67$\pm$0.27 & $-26.24$ & 4 \\[2pt]
J2336+1842 & 6.60$\pm$0.04 & $-1.22^{+0.19}_{-0.09}$ & 1.42$\pm$0.53 & 0.28$\pm$0.1 & 0.18$\pm$0.07 & 0.19$\pm$0.08 & -  & -  & - & $-24.32$ & 4 \\[2pt]
\hline \hline
\end{tabular}} \\
\vspace{0.2cm} \raggedright { \small \textbf{References.} [1] \cite{Fan2004AJ....128..515F} [2] \cite{Belladitta2020A&A...635L...7B} [3] \cite{Yang2021ApJ...923..262Y} [4] \cite{Gloudemans2022A&A...668A..27G} [5] \cite{WangF2016} } 
\end{table*}

We reduced the GMRT data using the software package Source Peeling and Atmospheric Modelling (SPAM; \citealt{Intema2014ASInC..13..469I, Intema2014ascl.soft08006I, Intema2017A&A...598A..78I}), which is based on the Astronomical Image Processing System (AIPS; \citealt{Greisen2003ASSL..285..109G}) python data reduction scripts and includes direction-dependent calibration and imaging. We followed the standard reduction procedure\footnote{\url{http://www.intema.nl/doku.php?id=huibintema:spam:pipeline}} and split the wideband data into 6 subbands before calibration. Each subband has a width of $\sim$33, 50, and 60 MHz for band 3, 4, and 5, respectively. If available, we made a first image with the narrowband data (band width 32 MHz) and used the Python Blob Detector and Source Finder (PyBDSF; \citealt{Mohan2015ascl.soft02007M}) to extract a source model. Subsequently, we used this source model for the wideband data calibration. For the band 3 and 4 observations we did not include subband 6 (483 and 824 MHz respectively) in the final image, because of the higher noise levels caused by decreased efficiency of the reflecting surface of the dishes at higher frequency. For the band 5 observations the root mean square (RMS) noise of the subband 6 image was similar to the other subbands and could therefore be included. Since our quasars are faint ($\sim1$ mJy), we only investigate the subband flux densities for the bright blazar J0309+2717 in Sect.~\ref{subsubsection:J0309+2717}. The flux density scale was set by SPAM following the prescriptions from \cite{Scaife2012MNRAS.423L..30S}. Finally, we combined all wideband calibrated images using WSClean \citep{Offringa2014MNRAS.444..606O} and extracted the resulting source flux densities using PyBDSF. The mean frequencies of our band 3, 4, and 5 images are 383, 675, and 1260 MHz, respectively. All quasars were detected in each of the GMRT bands. The beam sizes and average RMS of the resulting reduced images are summarised in Table ~\ref{tab:obs_details}. The final images are shown in Appendix \ref{sec:appendix_images}.  

\subsection{Archival radio data}

All quasars in our sample have been detected in the LOFAR Two-Metre Sky Survey second data release (LoTSS-DR2; \citealt{Shimwell2022A&A...659A...1S}) at 144 MHz, which covers over 5720 deg$^2$ of the Northern sky at 6$\arcsec$ resolution and with a median RMS noise level of $\sim83\ \mu$Jy beam$^{-1}$. To further increase our frequency coverage, we used radio images of our quasar sample from the VLA FIRST survey at 1.4 GHz \citep{Becker1994ASPC...61..165B} ($\sim$0.1 mJy noise level), the Very Large Array Sky Survey (VLASS; \citealt{Lacy2020PASP..132c5001L}) at 2-4 GHz ($\sim$0.12-0.17 mJy noise level), and the Rapid ASKAP Continuum Survey (RACS; \citealt{McConnell2020PASA...37...48M}) at 887 MHz. Four of our quasars are located within the FIRST footprint with 2 detections at $>5\sigma$ significance. VLASS covers our full sample with detections of 5 sources. The LoTSS-DR2, FIRST, and VLASS (epoch 2.1 and 2.2) images of our sources are also shown in Appendix \ref{sec:appendix_images}. For the LoTSS-DR2 survey and the FIRST survey, we adopted the measurements from the published catalogues. For the VLASS survey we have extracted the flux densities from the images using PyBDSF. Six of our quasars are covered by RACS at 887 MHz (and 25$\arcsec$ resolution) with three detections at $>5\sigma$. None of our sources are resolved in RACS and their flux densities have been determined using a single Gaussian fit with the Common Astronomy Software Applications (CASA; \citealt{Casa2022PASP..134k4501C}). We have added an uncertainty of 10\% of the flux density for each of the flux density errors for all four surveys to account for systematics between the different datasets. 

One of the quasars in our sample, J1133+4814, is located within the footprint of the LOFAR LBA Sky Survey (LoLSS; \citealt{deGasperin2023arXiv230112724D}) DR1 at 54 MHz and covering 650 deg$^2$ of the HETDEX spring field. However, this quasar has a neighbouring bright radio source at $\sim$27$\arcsec$ distance and the resolution of LoLSS of 15$\arcsec$ is not high enough to distinguish them.

\section{Spectral modelling}
\label{sec:modelling}

Since the main goal of this paper is to investigate the radio spectral energy distribution (SED) properties of these high redshift quasars, in this section we detail the models used for fitting. We consider both a simple power-law model and two types of absorption models (FFA and SSA), which have been adopted from \cite{callingham2015ApJ...809..168C} and are the two main models in explaining the turnover in the spectra of radio sources. A simple power-law can be produced by non-thermal synchrotron emission and is described by
\begin{equation}
    S_{\nu} = a \, \nu^{\alpha}, 
\end{equation}
with $\nu$ is the frequency, $\alpha$ the radio spectral index, and $a$ the amplitude of the synchrotron spectrum. 

\subsection{Free-Free Absorption}

FFA is caused by attenuation of emission by an internal or external ionised screen, which can either be homogeneous or inhomogeneous. Details of such models are outlined by \cite{callingham2015ApJ...809..168C}. These models assume that the radio plasma produces a non-thermal power-law spectrum. In the case of a homogeneous external screen, a foreground screen absorbs the free electrons and the resulting absorbed spectrum is given by 
\begin{equation} 
\label{eq:hom_ffa}
    S_{\nu} = a \, \nu^{\alpha}e^{-\tau_{\nu}},
\end{equation}
with $\tau_{\nu} = (\nu/\nu_{p})^{-2.1}$ the optical depth and $\nu_p$ the frequency at which $\tau_{\nu} = 1$ and the source becomes optically thick. In the case of an internal screen, the absorbing ionised medium is mixed with the electrons that are producing the non-thermal synchrotron emission. The spectrum can then be described as
\begin{equation} 
    S_{\nu} = a \, \nu^{\alpha} \, \Bigg( \frac{1-e^{-\tau_{\nu}}}{\tau_{\nu}} \Bigg).
\end{equation}

Finally, the homogeneous FFA can also show an exponential break due to an ageing electron population. This model is given by

\begin{equation}
    S_{\nu} = a \, \nu^{\alpha} \, e^{-\tau_{\nu}} \, e^{-\nu/\nu_{b}},
\end{equation}

with $\nu_b$ the break frequency. This formula is similar to Eq.~\ref{eq:hom_ffa} apart from the additional factor $e^{-\nu/\nu_{b}}$ that produces the high frequency exponential cutoff.

\subsection{Synchrotron Self-Absorption}

As explained by \cite{Kellermann1966AuJPh..19..195K}, in the case of SSA the turnover occurs when the source has a brightness temperature higher than the plasma of non-thermal electrons. The turnover frequency is then the frequency at which the electrons that emitted the photons have the highest chance to absorb them again. Assuming a homogeneous emitting region, the SSA model is given by
\begin{equation} 
    S_{\nu} = a \, \Bigg(\frac{\nu}{\nu_{p}}\Bigg)^{-(\beta-1)/2} \Bigg( \frac{1-e^{-\tau}}{\tau}\Bigg),
\end{equation}
\noindent with 
\begin{equation}
    \tau = \Bigg(\frac{\nu}{\nu_{p}}\Bigg)^{-(\beta+4)/2},
\end{equation}
where $\alpha = -(\beta-1)/2$. We refer the reader to \cite{callingham2015ApJ...809..168C} for a more detailed overview of these absorption models and their physical mechansims. \\

\begin{figure*}
    \centering
    \includegraphics[width=1.0\textwidth]{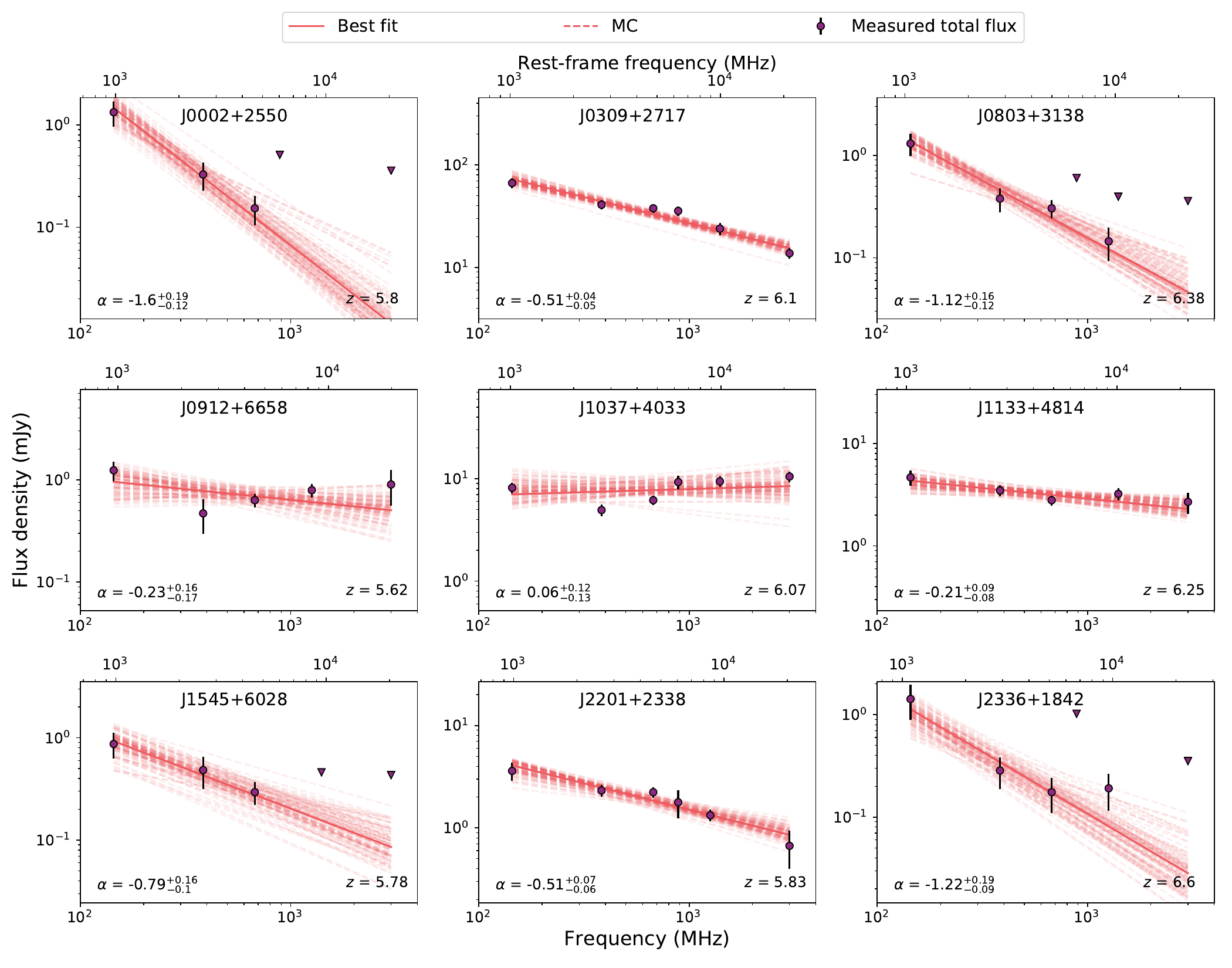}
    \caption{Radio spectra of our sample of 9 quasars observed with GMRT and including LoTSS-DR2, RACS, FIRST, and VLASS measurements. Non-detections in the FIRST and VLASS survey are given as 3$\sigma$ upper limits. Each spectrum is well described by a power law with varying spectral indices (indicated in the bottom left corner) and there is no evidence for a spectral turnover, except for J0309+2717 (see Sect.~\ref{subsubsection:J0309+2717}). The upper x-axis indicates the rest-frame frequency. The flat or negative slopes of these spectra suggest a turnover below 1 GHz in rest-frame.}
    \label{fig:spectra}
\end{figure*}

\subsection{Fitting procedure}
For the fitting procedure, we use an invariant Markov chain Monte Carlo (MCMC) approach as implemented by the \textsc{emcee} python package \citep{emcee2013PASP..125..306F}. To determine the best fit, we calculate the 50th percentile for each parameter (after 5000 iterations) and the 16th and 84th percentile for the errors. To compare the model fits, we use the Bayesian Information Criterion (BIC), which in the case of Gaussian distributed model errors is given by
\begin{equation}
    \text{BIC} = \chi^2 + k \ln N,
\end{equation}
with $\chi^2$ the chi-square statistic, $k$ the number of parameters, and $N$ the number of observations (see \citealt{kass1995bayes}). A difference between the BIC value of two model fits then gives either no evidence ($-2<\Delta$BIC$<2$), weak evidence ($|\Delta \text{BIC}| >2$), or strong evidence ($|\Delta \text{BIC}|> 6$) of one model being favoured over the other.

\section{Results}
\label{sec:results}

The resulting radio spectra are shown in Fig.~\ref{fig:spectra} with the purple points showing the integrated flux densities at their observed frequency with the rest-frame frequency indicated on the upper x-axis. Upper limits (3$\sigma$) are given for the quasars that remain non-detected in the FIRST, VLASS, and RACS survey. These upper limits on the flux densities are in all cases higher than the scatter around the power-law fit and therefore do not constrain the fit. All spectra are fitted well by a simple power-law. However, in the case of J0309+2717, we are potentially probing the onset of the spectral turnover, which is discussed in Sect.~\ref{subsubsection:J0309+2717}. The resulting MCMC fits are shown by the red solid line in Fig.~\ref{fig:spectra} together with 100 randomly selected MCMC fits (red dashed lines) to visualize the uncertainty in the fit. The spectral indices span a large range of $-1.6$ to 0.06 with only one source J1037+4033 having a positive spectral index, which is also consistent with zero. 

\begin{figure}
    \centering
    \includegraphics[width=1.0\columnwidth]{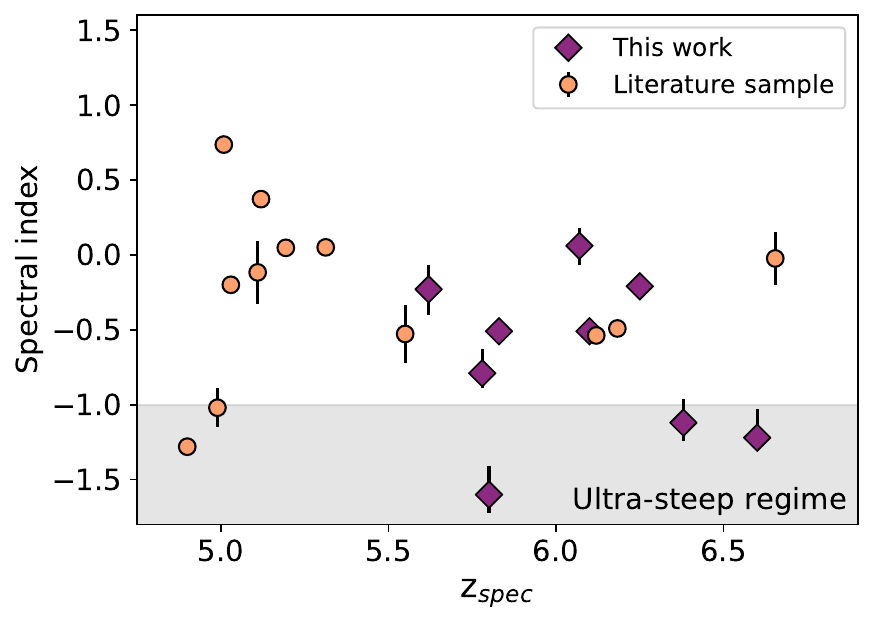}
    \caption{Radio spectral indices measured in this work versus spectroscopic redshift. The spectral indices of known quasars compiled by \cite{Gloudemans2021A&A...656A.137G, Gloudemans2022A&A...668A..27G} that are detected by LOFAR and either FIRST or VLASS are plotted for comparison. The sources from Fig.~\ref{fig:radio_lum} without radio spectral index measurements are excluded. Our sample covers a wide range of spectral indices and contains 3 sources that could be classified as ultra-steep spectrum sources ($\alpha<-1.0$).} 
    \label{fig:spec_index}
\end{figure}

We find 3 quasars (J0002+2550, J0803+3138, and J2336+1842) that could be classified as ultra-steep spectrum sources ($\alpha < -1$). A steep radio spectrum can be caused by extended radio lobe emission or radiative losses caused by for example ageing of the radio jet, which we will discuss further in Section ~\ref{subsec:discuss_radio_spec}. We also find 3 quasars with a flat spectral index (J0912+6658, J1037+4033, J1133+4814), which could be potential blazars (see Section~\ref{subsec:discuss_radio_spec}). The spectral indices of our sample and literature quasars are shown in Fig.~\ref{fig:spec_index}. Only a few quasars at $z>5$ have been discovered with ultra-steep spectra. The other two literature quasars with $\alpha < -1$ around $z\sim5$ were also discovered in \cite{Gloudemans2022A&A...668A..27G}. The steepest-spectrum radio source J0002+2550, with $\alpha = -1.6^{+0.19}_{-0.12}$, was initially discovered by \cite{Fan2004AJ....128..515F} in the Sloan Digital Sky Survey without any radio selection. However, it was selected for our follow-up observations because of its detection by LOFAR. 

We note that some of the flux density measurements deviate slightly from the power-law fit. These outliers are potentially caused by flux density calibration issues and beam size variations, however, these are expected to fall within the 10\% uncertainty taken on all of the flux densities. Variability could also play a role in the observed spectrum, since our observations are not taken simultaneous \citep[see e.g.][]{Orienti2010MNRAS.408.1075O}. Especially blazars, like J0309+2717, are known to be highly variable on short timescales (see Sect.~\ref{subsubsection:J0309+2717} and \ref{subsec:discuss_radio_spec}). The observed variability is generally less extreme at low-frequencies ($<1$ GHz; \citealt{Fan2007A&A...462..547F}), however, since we are studying high-$z$ sources, we are still probing the rest-frame GHz emission and therefore variability could have impacted the observed spectral shapes and created potential outliers from the power-law fit. Finally, the radio spectra could consist of multiple absorption components \citep[see e.g.][]{Shao2022A&A...659A.159S}, however, more high- and low-frequency observations are needed to investigate this. 

Since we do not detect the turnover, we cannot constrain the turnover peak with the absorption models discussed in Sect.~\ref{sec:modelling} for most of our sources. However, for J0309+2717 we can constrain the spectral shape with additional published data as we discuss in Sect.~\ref{subsubsection:J0309+2717}. The lack of peaked spectrum sources in our sample is interesting, since this is somewhat in contrast with earlier results found by \cite{Shao2022A&A...659A.159S} and other high-$z$ quasars, which will be further discussed in Sect.~\ref{subsec:discuss_radio_spec}. The power-law spectral indices, integrated flux densities, and derived M$_{1450\AA}$ values are given in Table~\ref{tab:measurements}. To improve readability, the complete radio measurements including peak flux densities are quoted in Appendix \ref{sec:appendix_images}. 

\begin{figure*}
    \centering
    \begin{subfigure}{\columnwidth}
        \centering
        \vspace{-0.2cm}
        \includegraphics[width=\textwidth, trim={0cm 0cm 0cm 0.0cm}, clip]{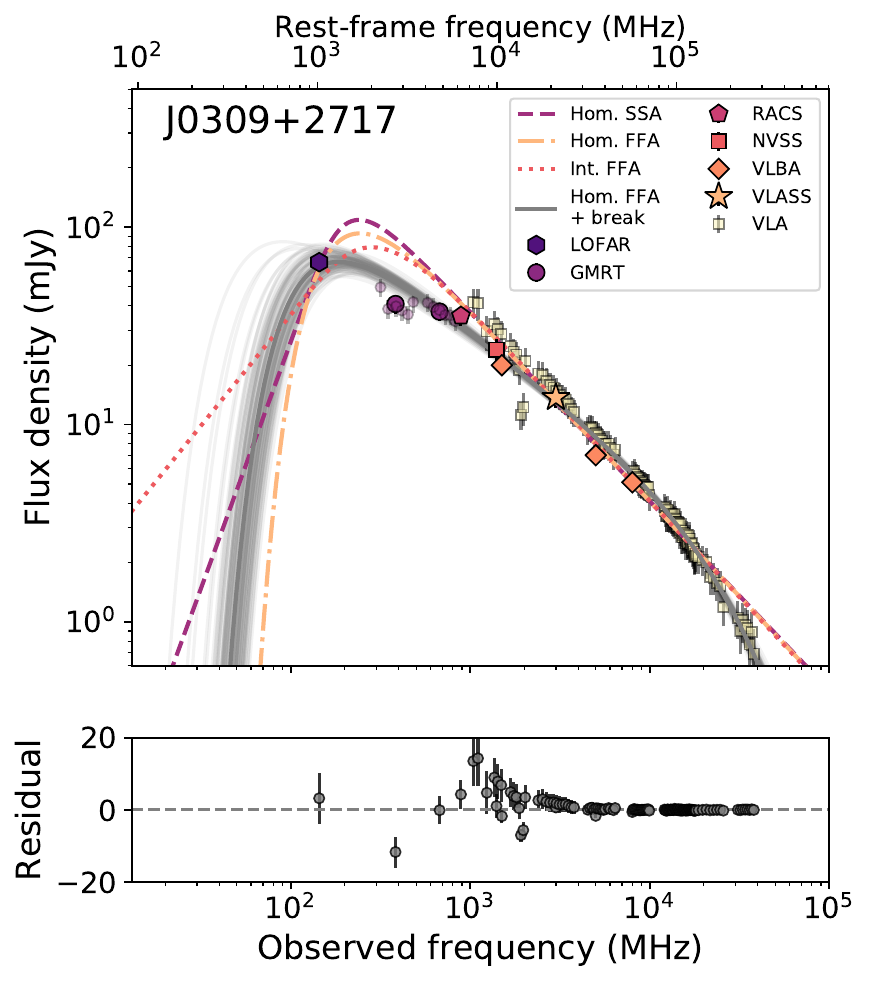}
    \end{subfigure}
     ~ 
    \begin{subfigure}{\columnwidth}
        \centering
        \vspace{-0.2cm}
        \includegraphics[width=\textwidth, trim={0cm 0cm 0cm 0.0cm}, clip]{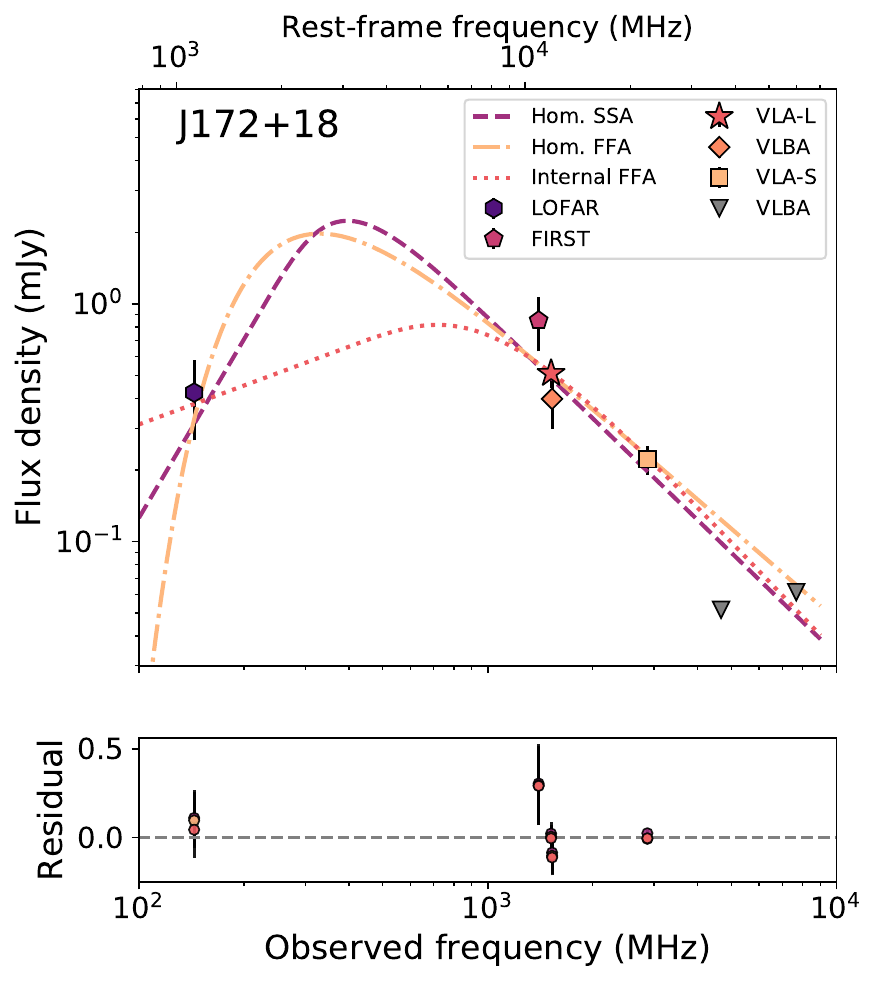}
    \end{subfigure}
    \caption{\label{fig:radio_curves} Radio spectra of J0309+2717 (left) and J172+18 (right) including reported literature radio flux density measurements. Left: The GMRT subband flux densities of the band 3 and 4 observations are shown in light purple. The VLBA and VLA measurements have been provided by \cite{Spingola2020A&A...643L..12S}. The dashed, dash-dotted and dotted line show the best fitting homogeneous SSA, homogeneous FFA, and internal FFA models, respectively. The homogeneous FFA with a break due to an ageing electron population (grey line) gives the best fit to the spectrum with a reduced $\chi^2$ of 0.6 and $\Delta$BIC $>6$ compared to all other models. Randomly selected MCMC fits are shown in light grey to visualize the uncertainty in the model fit with the parameters given in Tab.~\ref{tab:curve_parameters}. The bottom plot shows the residual of the homogeneous FFA model with a break. Right: The VLA-L and -S observations have been conducted by \cite{Banados2021ApJ...909...80B} and the VLBA observations by \cite{Momjian2021AJ....161..207M}. Upper limits are given for the LOFAR at 144 MHz and the VLBA observations at 4.67 and 7.67 GHz. Due to a limited number of data points between 150 MHz and 1 GHz the absorption models remain unconstrained. However, we can conclude that the spectrum likely peaks between 0.25-1.3 GHz (rest-frame 2-10 GHz), meaning it can be classified as a GPS source. Again the residuals are shown in the bottom panel. }
\end{figure*}

\subsection{Notes on individual sources}

Here we discuss some of the individual quasars in our sample that have been discovered and/or studied previously in other work. 

\subsubsection{J0002+2550}

This quasar at $z=5.80$ was discovered by \cite{Fan2004AJ....128..515F} from the Sloan Digital Sky Survey (SDSS; \citealt{York2000AJ....120.1579Y, Lyke2020ApJS..250....8L}) and its spectral energy density (SED) shape (from 0.1-80 $\mu$m) has been studied by \cite{Leipski2014ApJ...785..154L}, resulting in an upper limit on the star formation rate (SFR) of $<1.5\times 10^{3}$ M$_{\odot}$ yr$^{-1}$. Our GMRT observations show that it has a very steep radio spectrum of $\alpha=-1.6^{+0.19}_{-0.12}$. The possible origin of the steep spectrum will be further discussed in Sect.~\ref{subsec:discuss_radio_spec}.

\subsubsection{J0309+2717}
\label{subsubsection:J0309+2717}
J0309+2717 is a blazar at $z=6.1$ discovered by \cite{Belladitta2020A&A...635L...7B} and its radio, X-ray, and SMBH properties have been studied in detail previously (e.g. \citealt{Belladitta2020A&A...635L...7B, Spingola2020A&A...643L..12S, Belladitta2022A&A...660A..74B}, Spingola in prep.). The Very Long Baseline Array (VLBA) observations conducted by \cite{Spingola2020A&A...643L..12S} resolve the radio jet at milliarcsecond (mas) resolution, which reveals a one-sided jet with an extent of about 0.5 kpc and shows multiple sub-components. The radio spectrum presented by \cite{Spingola2020A&A...643L..12S} suggests a spectral index of $\alpha=-0.98\pm0.05$ between 1 and 40 GHz. By including the data points from 144 MHz, we find that the spectral index becomes flatter ($\alpha=-0.51\pm0.05$) at lower frequencies.

A potential low-frequency turnover was suggested by \cite{Spingola2020A&A...643L..12S}, who also observed a flattening of the spectrum between 147 MHz and 1.4 GHz. To further investigate the spectral shape we refitted the radio spectrum including the VLBA and VLA measurements from \cite{Spingola2020A&A...643L..12S} in the left panel of Fig.~\ref{fig:radio_curves} with the absorption models as described in Sect.~\ref{sec:modelling}. The homogeneous FFA with a break due to an ageing electron population gives the best fit to the spectrum with a reduced $\chi^2$ of 0.6 and $\Delta$BIC $>6$ with all other models (see Table \ref{tab:curve_parameters}). The best fitting model suggests the spectrum peaks at $110_{-20}^{+10}$ MHz observed frame, which translates to $750_{-110}^{+100}$ MHz rest frame. The high-frequency break is indicative of radiative losses (i.e. inverse Compton losses) from an ageing electron population in the radio lobes \citep[e.g.][]{Jaffe1973A&A....26..423J, Murgia2003PASA...20...19M, callingham2015ApJ...809..168C}. This high-frequency break was also found by \cite{Spingola2020A&A...643L..12S} at 14.5$\pm$0.5 GHz, however, they conclude their broken power-law model is not a significantly better fit to the data compared to their single power-law model. 

The excess flux around 1 GHz as measured in the VLA-L band could be due to variability since \cite{Spingola2020A&A...643L..12S} found an indication of 20-30\% variability between 1.4 and 3 GHz, which is a typical blazar characteristic. The measured flux density of our GMRT band-3 observation at 383 MHz is lower than expected from the model fits. To further investigate this we extract the flux densities from the six individual subband images as shown with light purple points in Fig.~\ref{fig:radio_curves} for both band 3 and 4. The integrated flux densities from band 4 follow the model curve, however, for band 3 this is not the case. While we have not been able to identify any problems due to the imaging process, we nevertheless expect it might be caused by systematics. Lower frequency observations at $<100$ MHz would help to pin down the exact emission mechanism (see Spingola et al in prep.). For now, we conclude the turnover peak likely occurs between a frequency of 90-330 MHz in observed frame and 0.6-2.3 GHz in rest-frame. 

Furthermore, \cite{Belladitta2022A&A...660A..74B} studied its near-infrared (NIR) spectrum using the Large Binocular Telescope and deduced a high SMBH mass of $1.45^{+1.89}_{-0.85} \times 10^9$ M$_{\odot}$, which challenges existing models of SMBH growth in the early Universe. 

\subsubsection{J0803+3138}
This quasar at $z=6.384$ has been discovered by \cite{Wang2019ApJ...884...30W} using the DESI Legacy Imaging survey and UKIRT Hemisphere survey. NIR spectroscopcic follow-up observations with Gemini/GNIRS by \cite{Yang2021ApJ...923..262Y} show that it also has a high SMBH mass of 1.40$\pm0.18 \times 10^9$ M$_{\odot}$ and M$_{1450\AA}$ of $-26.49$. According to our radio observations, this source can also be classified as steep spectrum with $\alpha = -1.12^{+0.16}_{-0.12}$. 

\subsubsection{J1037+4033}

J1037+4033 is the only quasar from our sample with a slightly positive (but consistent with zero) spectral index ($\alpha = 0.06^{+0.12}_{-0.13}$). Therefore, it could be turning over at $>3$ GHz (rest-frame >21 GHz) instead of below 144 MHz. This would make it a potential HFP source, which are generally thought to be young \citep[e.g.][]{Dallacasa2000A&A...363..887D}. More GHz-frequency observations (using for example the VLA or VLBA) or ultra-low frequency observations ($<50$ MHz) are necessary to confirm this. 

\subsubsection{J1545+6028}

This quasar at $z=5.78$ has been discovered by \cite{Wang2016ApJ...819...24W} using an SDSS-WISE quasar selection. They report a M$_{1450\AA}$ value of $−27.37$. No further NIR spectrum and derived SMBH mass have been published so far.

\subsubsection{J2336+1842}

This quasar at $z=6.6$ is part of the sample recently discovered by \cite{Gloudemans2022A&A...668A..27G} and is the highest redshift source in our GMRT observing campaign. The flux density of this quasar is close to the detection limit in GMRT band 4 and 5. According to our fitting routine it can be classified as a steep spectrum source with $\alpha=-1.22^{+0.19}_{-0.09}$, even though the band 5 integrated flux density is higher than expected. Deeper observations (RMS $<0.02$ mJy) in the GHz regime are needed to confirm the spectral slope. 

\section{Peaked spectrum quasar at z=6.8}
\label{subsec:banados_quasar}

One of the highest redshift radio-loud quasar to date is J172+18 at $z=6.82$, which was discovered by \cite{Banados2021ApJ...909...80B}. It has been observed previously in the VLA-S and -L band \citep{Banados2021ApJ...909...80B} and the VLBA at 1.53, 4.67, and 7.67 GHz \citep{Momjian2021AJ....161..207M}. The VLA-S and L-band observations show a steep negative spectral index of $\alpha=-1.31\pm0.08$, but the VLBA observations do not detect the quasar with a 3$\sigma$ upper limit of 52 and 61 $\mu$Jy at 4.67 and 7.67 GHz, respectively. As mentioned in \cite{Gloudemans2021A&A...656A.137G}, the measured flux density of this quasar is close to the detection limit of the LoTSS-DR2 survey at 144 MHz with a $\sim3\sigma$ tentative detection of 0.33$\pm$0.12 mJy beam$^{-1}$ and therefore J172+18 is likely a peak spectrum source. It has also been detected in the FIRST survey at 1.4 GHz with a flux density of 1.02$\pm$0.14 mJy, however, \cite{Banados2021ApJ...909...80B} measures 0.85$\pm$0.14 mJy, which we adopt in this work. 

The radio spectrum of J172+18 is shown in the right panel of Fig.~\ref{fig:radio_curves}. To constrain the spectral shape, we again fit the FFA and SSA models to the available flux measurements as described in Sect.~\ref{sec:modelling}. We correct the LOFAR peak flux measurement to a total flux density by multiplying by a factor of 1.3, which is the average correction for unresolved sources in LoTSS-DR2 (see \citealt{Gloudemans2021A&A...656A.137G}). There is no evidence that any of these models is favoured over the others ($|\Delta \text{BIC}| \leq 0.5$; see Table~\ref{tab:curve_parameters}). The spectrum does potentially flatten after the turnover, which has been observed before by \cite{Callingham2017ApJ...836..174C} for an ultra-steep spectrum source at $z=5.19$. The VLBA non-detection at 4.67 GHz can potentially be explained by radiative losses from an ageing electron population causing a break at high frequency similar to J0309+2717 (see \ref{subsubsection:J0309+2717}). Again, due to the limited number of radio detections, the model fits are not well constrained. Therefore, we can only conclude that the radio spectrum of J172+18 likely peaks between 0.25-1.3 GHz in the observed frame, which translates to 2-10 GHz in rest-frame. J172+18 can therefore be classified as a GPS source (as suggested also by \citealt{Momjian2021AJ....161..207M}). Future GMRT band-3 and -4 observations could further constrain the peak frequency.

\section{Discussion}
\label{sec:discussion}

Our GMRT observations of 9 radio-bright quasars at $z>5.6$ have led to the discovery of 3 ultra-steep spectrum sources, 3 flat-spectrum sources, 2 moderate steep spectrum sources, and 1 potential PS source. The lack of turnovers in these spectra indicates there is no strong SSA or FFA absorption in the observed frequency range. These results highlight the diversity of radio spectral shapes of the high-$z$ radio-loud quasar population. A detailed study of their radio properties will be crucial for the planning of future radio surveys (e.g. the SKA), to ensure maximising the number of detections and to be able to create automated methods to characterise all of these sources. In this section, we further discuss our findings, the lack of turnover spectra (apart from J0309+2717), and the linear size limits that can be placed based on the available observational data. 

\subsection{Radio spectral shapes}
\label{subsec:discuss_radio_spec}

Generally, the radio spectra of quasars consist of flat spectrum emission from the core and steep spectrum emission from the radio jets and lobes, if these exist, with absorption effects further altering the spectral shape. The flat radio spectrum from the core can be attributed to source inhomogeneity with different parts of the compact core region becoming optically thick due to SSA at different wavelengths and hence flattening the integrated radio spectrum. A steep spectrum will be produced by the radio jet and lobes which grow over time and potentially suffer from radiation losses when the active nucleus no longer supplies energetic electrons to the part of the jet where the radio emission originates from \citep[e.g.][]{Jaffe1973A&A....26..423J}. According to the unification model, the orientation of the AGN impacts whether the observed emission is dominated by this core or lobe emission. The absorption phase of an AGN, causing a turnover in the spectrum, is thought to be very short ($<$1 Myr) with the radio lobes of the PS sources observed to be small ($<20$ kpc). 

The fact that quasars are generally observed to have compact morphologies ($<10$ kpc) and flat radio spectra ($\alpha\sim-0.3$), whereas radio galaxies have more extended radio lobes ($>20$ kpc) and steep spectra ($\alpha$ < $-0.5$) (see e.g. \citealt{Kameno1995PASJ...47..711K, Miley2008A&ARv..15...67M, Tadhunter2016A&ARv..24...10T, Saxena2017MNRAS.469.4083S}), can be explained as an orientation and selection effect. In this work, our sample is selected to be bright at low-frequencies and therefore we are biased towards steep spectrum sources and against sources with a turnover. High-frequency studies on the other hand are biased towards flat spectrum sources and therefore combining different selections is key to probe the overall population of radio quasars. There is a subclass of steep spectrum radio quasars (SSRQs) that tend to be lobe-dominated \citep[e.g.][]{Urry1995PASP..107..803U, Athreya1997MNRAS.289..525A, Liu2006ApJ...637..669L}. Our 3 discovered ultra-steep spectrum sources are therefore likely also dominated by lobe emission with potentially radiative losses from aged material further steepening the spectrum. 

The three flat-spectrum sources we find (J0912+6658, J1037+4033, J1133+4814) could potentially be blazars (like J0309+2717), which have their relativistic jets oriented along the line of sight, making them extremely bright sources \citep[e.g.][]{Urry1995PASP..107..803U}. Because of their specific jet orientation, blazars can be used to efficiently trace the space density of the quasar population over cosmic time, which makes their identification useful \citep[e.g.][]{Ghisellini2010MNRAS.402..497G}.  Blazars are furthermore known to be highly variable in nature on timescales of hours to days \citep[e.g.][]{Ulrich1997ARA&A..35..445U}, which could again be problematic for constraining their radio spectra in this work as mentioned in Sect.~\ref{sec:results}. Especially J1037+4033 and J1133+4814 are exceptionally radio-loud with $R$-values of 1100$\pm$300 and 360$\pm$100, respectively (see \citealt{Gloudemans2022A&A...668A..27G}). J1133+4814 (also called P173+48) has also been identified as a blazar candidate by \cite{Banados2023ApJS..265...29B}, who find both a flat spectrum and variability. Follow-up radio monitoring and X-ray observations are necessary to confirm the blazar nature of these quasars.   

A previous study from \cite{Shao2022A&A...659A.159S} found 9 quasars at $5.0 \leq z \leq 6.1$ with spectral turnovers in the GHz regime. At lower redshift  \cite{Sotnikova2021MNRAS.508.2798S} classified 46\% of their sample of 102 quasars at $3 \leq z \leq 5$ as peaked-spectrum sources between 1.2-22 GHz and \cite{Krezinger2022ApJS..260...49K} found 2 peaked spectrum sources in their sample of 13 quasars at $4.05 \leq z \leq 4.47$ between 0.15-10.6 GHz. As discussed in Sect. ~\ref{subsec:banados_quasar}, there is evidence for a turnover in the radio spectrum of J172+18 and potentially J0309+27. However, we do not find any evidence for a turnover for the other 8/9 quasars in our sample in a similar frequency range. In the `young' scenario the absorption phase is thought to be only short-lived, therefore our lack of turnovers could suggest the AGN in our quasars have been on for a longer period of time. This would suggest large linear radio jet sizes, which we investigate further in Sect.~\ref{subsec:discuss_linear_sizes}. The `frustration' hypothesis suggests that the spectral turnover is caused by a dense environment, with a lower turnover frequency suggesting a lower density environment. Therefore, our quasars could be situated in low density environments, causing a lower frequency turnover. However, we emphasize, that by selecting bright sources at low-frequencies we are biased against turnover sources and with this data we cannot constrain the turnover frequency and absorption models. 

Follow-up ultra-low frequency observations will be necessary to further constrain the turnover frequencies of the sample in this work. The LoLSS survey at 54 MHz (378 MHz at $z=6$) and $\sim$1 mJy beam$^{-1}$ noise level should be able to detect or set a 3$\sigma$ limit on the turnover for 7/9 of our quasars, assuming the measured spectral indices. The LOFAR Decameter Sky Survey (LoDeSS; Groeneveld et al. in prep) will go even lower in frequency with $\sim$10 mJy beam$^{-1}$ noise level at 22 MHz (154 MHz at $z=6$), however, since our sources are generally faint, the LoDeSS survey will only be able to further constrain the spectral shape of J0309+2717.

\begin{table}
\caption{Deconvolved 3$\sigma$ size upper limits (except for J0309+2717) obtained using the highest resolution available observations with a significant source detection.}
\label{tab:sizes}      
\centering
\resizebox{1.0\columnwidth}{!}{
\begin{tabular}{c c c c} 
\hline\hline
Source name & DC 3$\sigma$ uplim & Survey & Size limit \\ 
 & (arcsec) & & (kpc)\\ 
\hline
J0002+2550 & 3.58 $\times$ 0.45 & GMRT-4 & 21 \\
J0309+2717 & 0.0128 $\times$ 0.0076 & VLBA 1.5 GHz $^1$ & 0.5 \\
J0803+3138 & 1.6 $\times$ 1.7 & GMRT-5 & 9  \\
J0912+6658 & 2.15 $\times$ 1.05 & VLASS & 13 \\
J1037+4033 & 1.16 $\times$ 0.4 & VLASS & 7  \\
J1133+4814 & 3.91 $\times$ 2.18 & VLASS & 22 \\
J1545+6028 & 4.45 $\times$ 2.73 & GMRT-4 & 26 \\
J2201+2338 & 1.58 $\times$ 1.09 & VLASS & 9 \\
J2336+1842 & 2.43 $\times$ 2.0 & GMRT-5 & 13 \\
\hline \hline
\end{tabular}}
\\ \vspace{0.1cm} \raggedright {\tiny \textbf{Notes.} [1] VLBA measurements from \cite{Spingola2020A&A...643L..12S}}
\end{table}

\begin{figure}
    \centering
    \includegraphics[width=1.0\columnwidth]{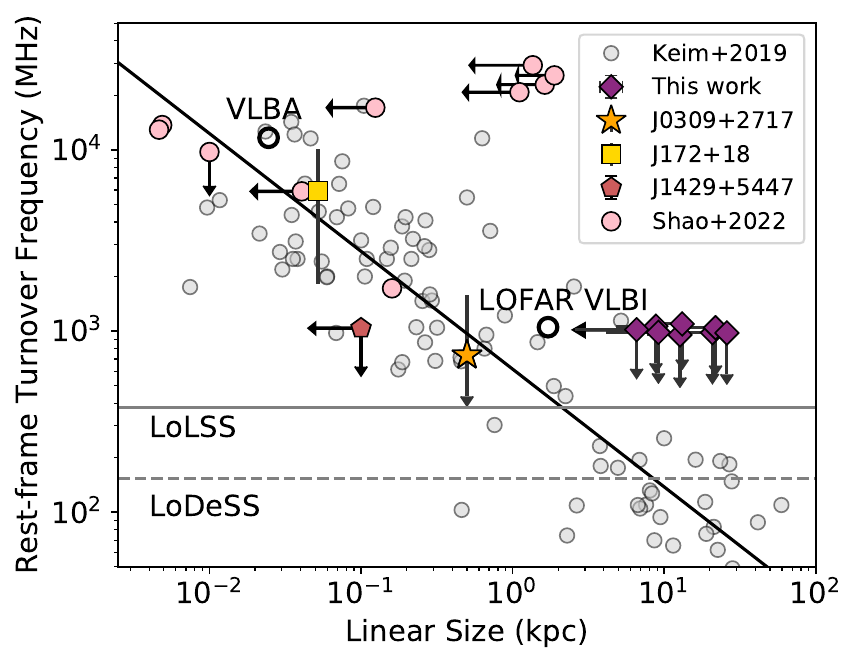}
    \caption{Linear sizes versus rest-frame turnover frequency for the quasars in our sample, J172+18, and the 10 high-$z$ quasars from \cite{Shao2022A&A...659A.159S}. The grey points show low-redshift ($z\approx0.1$-2) peaked spectrum sources from \cite{Keim2019A&A...628A..56K}, which follow a tight relation given by the black solid line (see \citealt{Keim2019A&A...628A..56K}). Apart from J0309+2717, we have only upper limits on the linear sizes of the quasars in our sample and their turnover frequencies. Five of the quasars from \cite{Shao2022A&A...659A.159S} have been observed at mas resolution with the VLBA and three of them have size measurements, which seem to follow the low-$z$ relation. The solid and dashed grey lines show the frequencies of the LoLSS and LoDeSS survey, respectively, and the black open circles the resolution of the LOFAR VLBI and VLBA at 150 MHz and 1.6 GHz (observed frame).} 
    \label{fig:linear_sizes}
\end{figure}

\subsection{Linear sizes}
\label{subsec:discuss_linear_sizes}

All of our sources are unresolved in the LOFAR, GMRT, and VLA observations and therefore we can only place upper limits on their physical sizes (except for J0309+2717, which has been observed with the VLBA). For each quasar we use the highest frequency radio detection and the deconvolved FWHM of the major axis of the source as determined by PyBDSF and their uncertainties to get an 3$\sigma$ upper limit on their sizes. The derived size upper limits are summarised in Table \ref{tab:sizes}. As discussed in Sect.~\ref{subsubsection:J0309+2717}, the resolved jet properties of J0309+2717 have been investigated by \cite{Spingola2020A&A...643L..12S} and they find this quasar has a bright one-sided jet extended for $\sim$0.5 kpc. Furthermore, as discussed in Sect.~\ref{subsec:banados_quasar}, J172+18 has been observed with the VLBA by \cite{Momjian2021AJ....161..207M}, which resulted in a linear size measurement of 52 pc. The upper limits on the sizes of the quasars in our sample and the rest-frame turnover frequencies (144 MHz shifted to rest-frame) have been plotted in Fig.~\ref{fig:linear_sizes}, together with the measurements of J0309+2717 and J172+18. We note that the VLBA observations potentially resolve out diffuse radio lobe emission, in which case the size measurement is of the inner jet, which cannot be directly compared to extended radio lobe sizes from our low-frequency observations. At low-redshift ($z\approx0.1-2$) a relation between the linear size and turnover frequency has been established by \cite{Keim2019A&A...628A..56K} and \cite{Orienti2014MNRAS.438..463O}. We plot this relation also in Fig.~\ref{fig:linear_sizes} as given by \cite{Keim2019A&A...628A..56K}.

We also plot the 10 quasars between $5.0 \leq z \leq 6.1$ studied by \cite{Shao2022A&A...659A.159S} including J1427+3312, which is also very bright at low frequencies (see Sect.~\ref{subsec:sample_selection}). Five of them have been observed previously at mas resolution (see \citealt{Momjian2003AAS...203.7813M, Frey2005A&A...436L..13F, Momjian2008AJ....136..344M, Frey2010A&A...524A..83F, Gabanyi2015MNRAS.450L..57G}) with three of them having resolved size measurements. For the other 5 quasars, we used the beam size from the VLA X-band observations from \cite{Shao2022A&A...659A.159S} as an upper limit on the linear size. Their results show that an inhomogeneous FFA model can accurately describe their observed spectra and therefore we use those peak frequency measurements in Fig.~\ref{fig:linear_sizes}. Only one quasar in their work did not show direct evidence for a turnover (J2228+0110) and this source is plotted with an upper limit for the turnover frequency ($\leq10$ GHz rest frame). Its linear size of $\sim$10 pc is adopted from \cite{Cao2014A&A...563A.111C}, who conducted European VLBI Network observations. We also plot the upper limits for J1429+5447 at $z=6.2$, which has been studied by \cite{Shao2020A&A...641A..85S} and \cite{Frey2011A&A...531L...5F}. Similar to J1427+3312, this quasar is bright at low frequency, but did not show evidence of a turnover between 120 MHz - 5 GHz observed frame in \cite{Shao2020A&A...641A..85S} and its linear size has been determined to be $<100$ pc by \cite{Frey2011A&A...531L...5F}. Besides J1429+5447, the few measurements of high-$z$ quasars seem to be in agreement with the established relation at low-$z$. Finally, we also indicate the turnover frequencies probed by the LoLSS and LoDeSS survey at $z=6$. 

If the relation from \cite{Keim2019A&A...628A..56K} holds at high-$z$, it would suggest the quasars in our sample are expected to have relatively large linear sizes ($\sim$0.5-30 kpc) and low turnover frequencies ($\sim$0.1-1 GHz rest-frame). However, with current observations we are unable to confirm the large linear sizes and low frequency turnovers. Future follow-up high-resolution radio observations are needed to constrain the sizes, using for example the VLBA which can reach mas resolution at GHz frequencies (e.g. $\sim$10 mas at 1.6 GHz; 57 pc at $z=6$). At lower frequencies, LOFAR VLBI can achieve a resolution of 0.3$\arcsec$ at 150 MHz ($\sim$1.7 kpc at $z=6$; see Fig. ~\ref{fig:linear_sizes}).

\section{Conclusion}
\label{sec:conclusion}

In this work, we have conducted GMRT observations of a sample of 9 of the highest redshift quasars that are bright at low-frequencies to model their radio spectra and constrain their emission mechanism. We did not find any evidence of a spectral turnover in our sample, except for J0309+2717, which implies we do not detect any strong FFA or SSA absorption in their spectra. The spectrum of J0309+2717, including additional VLBA data, shows a potential turnover at 0.6-2.3 GHz rest-frame with a high frequency spectral break, which could be indicative of radiative ageing of the electron population in the radio lobes. The spectral indices of our sources are in a wide range and vary from $-1.6$-0.06. We identify three quasars with ultra-steep spectra ($\alpha < -1$), which could be due to their orientation (making them radio lobe dominated) or radiative losses by for example ageing of the radio jet. Two of these ultra-steep spectrum quasars have been discovered in purely optical and near-infrared surveys, but one of them required a LOFAR detection in the selection process (J2336+1842 at $z=6.6$), thereby biasing the sample towards brighter objects at low frequencies. 

Furthermore, we also model the spectrum of J172+18 at $z=6.8$, which is only tentatively detected by LOFAR at 144 MHz and shows evidence for a turnover between 2-10 GHz rest-frame according to our model fitting and can therefore be classified as a GPS source. To investigate the relation between linear size and rest-frame turnover frequency, we determine upper limits on the sizes of the radio jets and turnover frequencies, with J0309+2717 having an extended jet of $\sim$500 pc as measured by \cite{Spingola2020A&A...643L..12S}. Compared to other known quasars at $z>5$, the quasars in our sample are expected to have lower turnover frequencies ($\lesssim1$ GHz rest-frame). If the relation found previously at low-redshift holds, this would indicate that the quasars in our sample have relatively extended radio jets ($\sim$0.5-30 kpc). Future ultra-low frequency observations ($<50$ MHz) with for example LOFAR and high-resolution follow-up observations with for example the VLBA are necessary to measure their radio jet sizes and constrain their spectral turnover, which will also allow for us to distinguish among various radio emission and absorption mechanisms. This work highlights the diversity of the radio spectral shapes of the high-$z$ quasar population. Further characterising them using the next-generation radio interferometers will be important for the planning of radio surveys, as well as understanding the physics of radio emission from quasars in the very early Universe.

\begin{acknowledgements}
{
KJD acknowledges funding from the European Union's Horizon 2020 research and innovation programme under the Marie Sk\l{}odowska-Curie grant agreement No. 892117 (HIZRAD) and support from the STFC through an Ernest Rutherford Fellowship (grant number ST/W003120/1). CS acknowledges the support from the INAF grants 1.05.12.04.04. MJH acknowledges support from the UK STFC [ST/V000624/1].

This paper is based (in part) on data obtained with the International LOFAR Telescope (ILT) under project codes LC0 015, LC2 024, LC2 038, LC3 008, LC4 008, LC4 034 and LT10 01. LOFAR \citep{vanHaarlem2013A&A...556A...2V} is the Low Frequency Array designed and constructed by ASTRON. It has observing, data processing, and data storage facilities in several countries, which are owned by various parties (each with their own funding sources), and which are collectively operated by the ILT foundation under a joint scientific policy. The ILT resources have benefited from the following recent major funding sources: CNRS-INSU, Observatoire de Paris and Universit\'e d'Orl\'eans, France; BMBF, MIWF-NRW, MPG, Germany; Science Foundation Ireland (SFI), Department of Business, Enterprise and Innovation (DBEI), Ireland; NWO, The Netherlands; The Science and Technology Facilities Council, UK; Ministry of Science and Higher Education, Poland.

This research made use of the Dutch national e-infrastructure with support of the SURF Cooperative (e-infra 180169) and the LOFAR e-infra group. The J\"ulich LOFAR Long Term Archive and the German LOFAR network are both coordinated and operated by the J\"ulich Supercomputing Centre (JSC), and computing resources on the supercomputer JUWELS at JSC were provided by the Gauss Centre for Supercomputing e.V. (grant CHTB00) through the John von Neumann Institute for Computing (NIC).

This research made use of the University of Hertfordshire high-performance computing facility and the LOFAR-UK computing facility located at the University of Hertfordshire and supported by STFC [ST/P000096/1], and of the Italian LOFAR IT computing infrastructure supported and operated by INAF, and by the Physics Department of Turin university (under an agreement with Consorzio Interuniversitario per la Fisica Spaziale) at the C3S Supercomputing Centre, Italy.

}
	
\end{acknowledgements}

\bibliographystyle{aa}
\bibliography{bibliography.bib}

\begin{appendix}

\section{Radio images \& flux densities}
\label{sec:appendix_images}

\begin{sidewaysfigure*}
    \centering\includegraphics[width=0.9\textwidth]{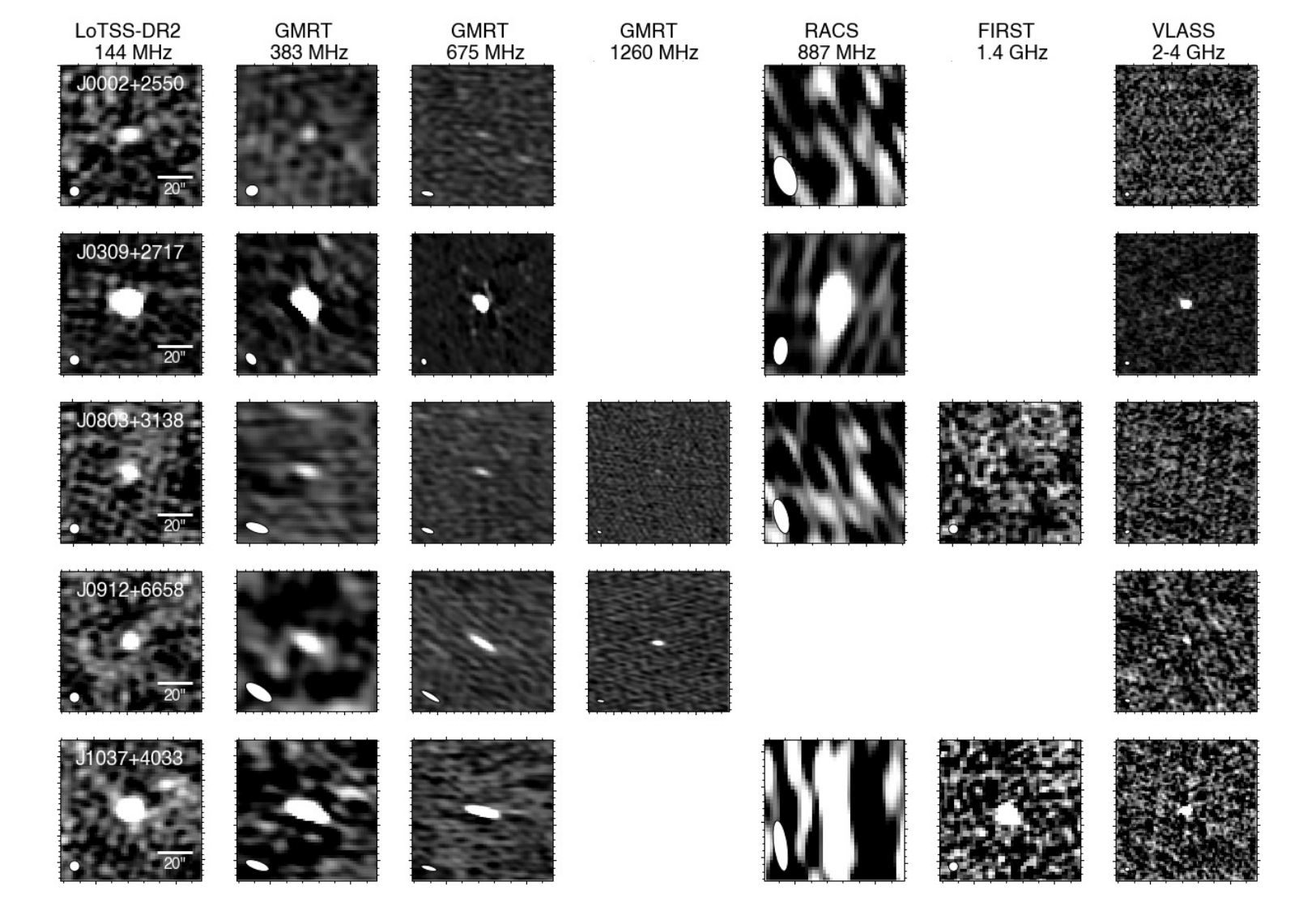}
    \caption{Cutouts (80$\arcsec \times$80$\arcsec$) of all quasars in our sample of LoTSS-DR2, GMRT, FIRST, and VLASS. The radio beam is shown in the bottom left corner of each panel. None of our quasars are resolved in these images. For the image scaling, the minimum and maximum data value (vmin and vmax) are fixed to $-1$ and 5 times the RMS noise, respectively, except for J0309+2717 and J1133+4814, where vmin and vmax are $-1$ and 10 times the RMS noise.}
    \label{fig:cutouts}
\end{sidewaysfigure*}

\begin{sidewaysfigure*}
    \centering\includegraphics[width=0.9\textwidth]{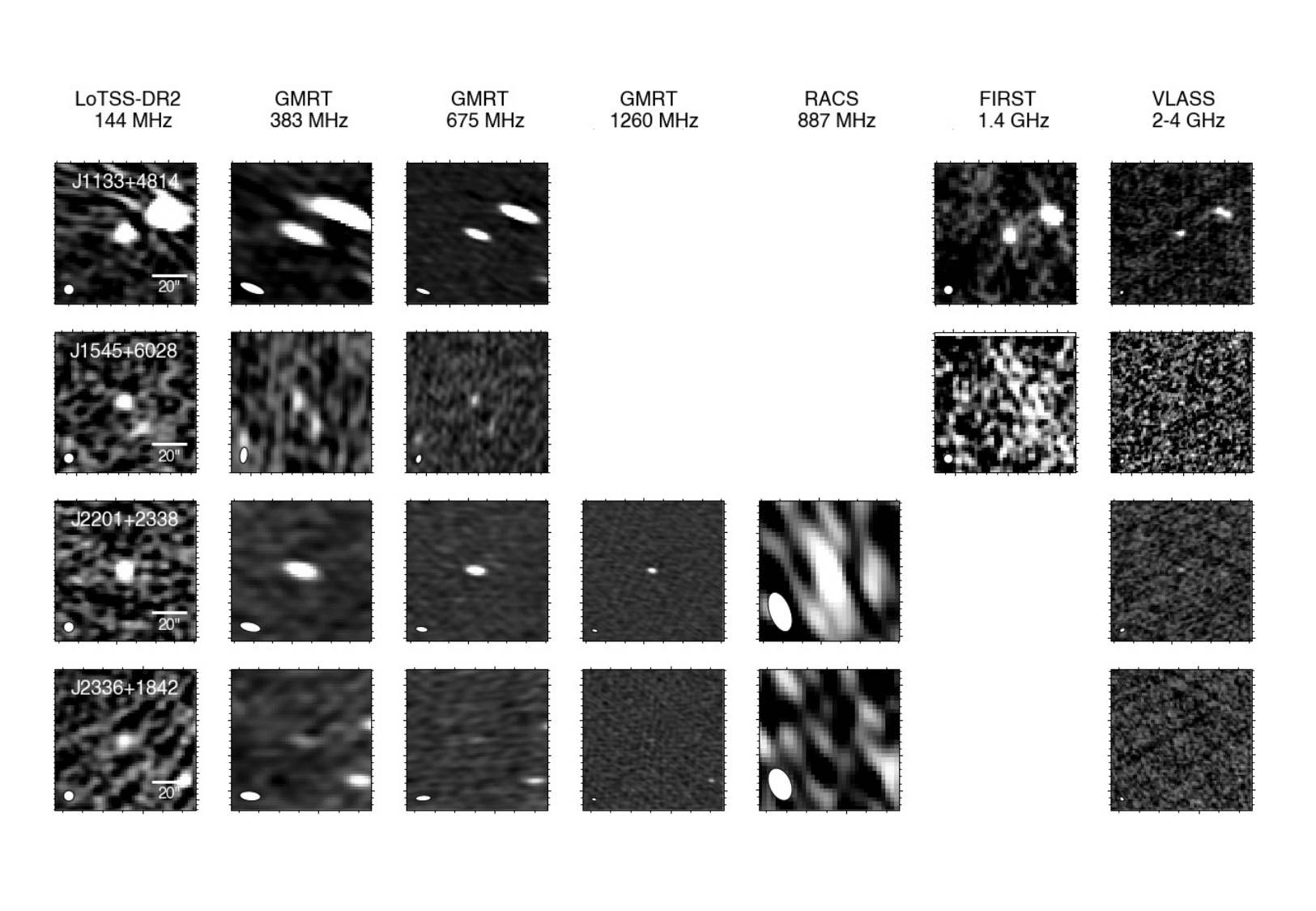}
    \caption{Same as Fig. ~\ref{fig:cutouts}}
    \label{fig:cutouts2}
\end{sidewaysfigure*}

Figure \ref{fig:cutouts} and \ref{fig:cutouts2} show the LoTSS-DR2, GMRT, RACS, FIRST, and VLASS images for each of the 9 quasars in our sample. All quasars have been detected in LoTSS-DR2 and our GMRT follow-up observations, however, they are not all detected in FIRST and VLASS. The extracted integrated radio flux densities and peak brightness are given in Table \ref{tab:radio_fluxes}.

\begin{table}[!bp]
  \centering

  \rotatebox{90}{%
    \begin{minipage}{\textheight}
      \def\arraystretch{2.0}
      \small\centering
      \setlength{\tabcolsep}{1.5mm}
      \scriptsize
      \centering
      \caption{Measured integrated radio flux densities and peak brightness from LoTSS-DR2, GMRT, RACS, FIRST, and VLASS. The integrated radio flux densities are in units of mJy and the peak radio flux densities in mJy beam$^{-1}$. }
      \label{tab:radio_fluxes}  
      \begin{tabular}{c*{15}{r}}
        \hline\hline
        Source name & S$_{144\text{MHz}}$ & S$_{144\text{MHz, peak}}$ & S$_{383\text{MHz}}$ & S$_{383\text{MHz, peak}}$ & S$_{675\text{MHz}}$ & S$_{675\text{MHz, peak}}$ & S$_{1260\text{MHz}}$ & S$_{1260\text{MHz, peak}}$ & S$_{887\text{MHz}}$ & S$_{887\text{MHz, peak}}$ & S$_{1.4\text{GHz}}$ & S$_{1.4\text{GHz, peak}}$ & S$_{3\text{GHz}}$ & S$_{3\text{GHz, peak}}$\\ 
        \hline
        J0002+2550 & 1.33$\pm$0.37 & 0.83$\pm$0.18 & 0.33$\pm$0.1 & 0.35$\pm$0.07 & 0.15$\pm$0.05 & 0.19$\pm$0.04 & - & -  & - & -  & - & -  & - & -  \\
        J0309+2717 & 66.52$\pm$7.05 & 54.05$\pm$5.52 & 40.83$\pm$4.3 & 38.47$\pm$3.91 & 37.73$\pm$3.85 & 35.38$\pm$3.56 & - & -  & 35.57$\pm$3.95 & 32.06$\pm$3.41 & 23.89$\pm$3.26 & 23.89$\pm$3.26 & 13.71$\pm$1.58 & 12.78$\pm$1.39 \\
        J0803+3138 & 1.3$\pm$0.32 & 0.97$\pm$0.19 & 0.38$\pm$0.1 & 0.43$\pm$0.08 & 0.3$\pm$0.06 & 0.3$\pm$0.05 & 0.14$\pm$0.05 & 0.15$\pm$0.04 & - & -  & - & -  & - & -  \\
        J0912+6658 & 1.23$\pm$0.26 & 1.04$\pm$0.18 & 0.47$\pm$0.17 & 0.56$\pm$0.13 & 0.63$\pm$0.1 & 0.63$\pm$0.08 & 0.79$\pm$0.12 & 0.76$\pm$0.1 & - & -  & - & -  & 0.9$\pm$0.34 & 0.76$\pm$0.2 \\
        J1037+4033 & 8.16$\pm$0.97 & 6.25$\pm$0.7 & 4.94$\pm$0.63 & 5.38$\pm$0.61 & 6.17$\pm$0.67 & 6.48$\pm$0.67 & - & -  & 9.28$\pm$1.42 & 8.44$\pm$1.07 & 9.42$\pm$1.09 & 9.28$\pm$1.08 & 10.51$\pm$1.28 & 9.87$\pm$1.11 \\
        J1133+4814 & 4.67$\pm$0.79 & 3.16$\pm$0.46 & 3.48$\pm$0.44 & 3.53$\pm$0.4 & 2.79$\pm$0.31 & 2.89$\pm$0.31 & - & -  & - & -  & 3.23$\pm$0.46 & 2.4$\pm$0.38 & 2.69$\pm$0.63 & 1.67$\pm$0.31 \\
        J1545+6028 & 0.87$\pm$0.24 & 0.6$\pm$0.13 & 0.48$\pm$0.17 & 0.31$\pm$0.08 & 0.29$\pm$0.07 & 0.23$\pm$0.04 & - & -  & - & -  & - & -  & - & -  \\
        J2201+2338 & 3.6$\pm$0.73 & 2.92$\pm$0.48 & 2.32$\pm$0.3 & 2.32$\pm$0.27 & 2.23$\pm$0.26 & 2.26$\pm$0.25 & 1.33$\pm$0.17 & 1.28$\pm$0.15 & 1.78$\pm$0.55 & 2.14$\pm$0.45 & - & -  & 0.67$\pm$0.27 & 0.67$\pm$0.18 \\
        J2336+1842 & 1.42$\pm$0.53 & 0.93$\pm$0.26 & 0.28$\pm$0.1 & 0.32$\pm$0.07 & 0.18$\pm$0.07 & 0.2$\pm$0.04 & 0.19$\pm$0.08 & 0.16$\pm$0.04 & - & -  & - & -  & - & -  \\
        \hline \hline        

      \end{tabular}
    \end{minipage}}
\end{table}

\section{Spectral modelling}
\label{appendix:curved_fits}

As discussed in Sect.~\ref{subsubsection:J0309+2717} and \ref{subsec:banados_quasar}, we fitted FFA and SSA absorption models to J0309+2717 and J172+18. The resulting fits are shown in Fig.~\ref{fig:radio_curves} and the spectral modeling parameters are given in Table \ref{tab:curve_parameters}.

\begin{table*}
\caption{Spectral model fitting parameters of J0309+2717 and J172+18. The model and fitting routine are specified in Sect.~\ref{sec:modelling}. The parameters are: the spectral index $\alpha$ of the synchrotron spectrum, the amplitude $a$ of the initial synchrotron spectrum, the peak frequency $\nu_p$ in the observed frame, the power law index of the electron energy distribution $\beta$, the break frequency $\nu_b$, and the Bayesian Information Criterion for each model fit.}
\label{tab:curve_parameters}      
\centering
\resizebox{0.85\textwidth}{!}{
\begin{tabular}{c c c c c c c c} 
\hline \hline
 & $\alpha$ & a & $\nu_{p}$ (MHz) & $\beta$ & $\nu_b$ (MHz) & BIC  \\
 \hline \hline
J0309+2717  &  &  &  &  &  &  \\
\hline
& \\[\dimexpr-\normalbaselineskip+2pt]
Homogeneous FFA  & $-0.97^{+0.02}_{-0.02}$ & 30000$^{+6300}_{-5000}$ & 170$^{+10}_{-10}$ &  &  & 220 \\
Homogeneous SSA  &  & 170$^{+10}_{-10}$ & 210$^{+10}_{-10}$ & 2.92$^{+0.04}_{-0.04}$ &  & 260 \\
Internal FFA  & $-0.98^{+0.02}_{-0.02}$ & 33000$^{+6900}_{-5700}$ & 300$^{+30}_{-30}$ &  &  & 190 \\
Homogeneous FFA + break & $-0.67^{+0.03}_{-0.03}$ & 3000$^{+810}_{-660}$ & 110$^{+10}_{-20}$ &  & 28000$^{+3800}_{-3100}$ & 89 \\
\hline
J172+18 &  &  &  &  &  & \\
& \\[\dimexpr-\normalbaselineskip+2pt]
\hline
Homogeneous FFA  & $-1.27^{+0.10}_{-0.05}$ & 5900$^{+2800}_{-3100}$ & 260$^{+200}_{-20}$ &  &  & 8.2 \\
Homogeneous SSA  &  & 3.5$^{+1.4}_{-1.3}$ & 380$^{+90}_{-60}$ & 3.8$^{+0.6}_{-0.5}$ &  & 8.4 \\
Internal FFA  & $-1.55^{+0.11}_{-0.05}$ & 57000$^{+30000}_{-34000}$ & 1100$^{+290}_{-210}$ &  &  & 7.9 \\
\hline \hline
\end{tabular}}
\\ 
\end{table*}

\end{appendix}

\end{document}